\documentclass[linenumbers]{aastex631}

\newcounter{ionctr}
\ifx \undefined \ion \newcommand{\ion}[2]{\setcounter{ionctr}{#2}{#1$\;${\small\rmfamily\Roman{ionctr}}\relax}} \fi

\begin{document}

\title{Inferring the CO$_2$ Abundance in Comet 45P/Honda-Mrkos-Pajdušáková from [\ion{O}{1}] Observations: Implications for the Source of Icy Grains in Cometary Comae}

\author[0009-0000-2828-2263]{Mikayla R. Huffman}
\affiliation{Laboratory for Atmospheric and Space Physics, \\
University of Colorado, \\
1234 Innovation Drive, \\
Boulder, CO 80303, USA \\
mikayla.huffman@colorado.edu} 

\author[0000-0002-0622-2400]{Adam J. McKay}
\affiliation{Appalachian State University, \\
525 Rivers St., \\
Boone, NC 28608, USA \\
mckayaj@appstate.edu} 

\author[0000-0003-4828-7787]{Anita L. Cochran}
\affiliation{University of Texas Austin/McDonald Observatory,\\
2512 Speedway, Stop C1402, \\
Austin, TX 78712, USA}

\begin{abstract}

The study of cometary composition is important for understanding our solar system’s early evolutionary processes. Carbon dioxide (CO$_2$) is a common hypervolatile in comets that can drive activity but is more difficult to study than other hypervolatiles due to severe telluric absorption. CO$_2$ can only be directly observed from space-borne assets. Therefore, a proxy is needed to measure CO$_2$ abundances in comets using ground-based observations. The flux ratio of the [\ion{O}{1}] 5577 Å line to the sum of the [\ion{O}{1}] 6300 Å and [\ion{O}{1}] 6364 Å lines (hereafter referred to as the [\ion{O}{1}] line ratio) has, with some success, been used in the past as such a proxy. We present an [\ion{O}{1}] line ratio analysis of comet 45P/Honda–Mrkos–Pajdušáková (HMP), using data obtained with the Tull Coudé Spectrograph on the 2.7-meter Harlan J. Smith telescope at McDonald Observatory, taken from UT February 21-23, 2017 when the comet was at heliocentric distances of 1.12-1.15 AU. HMP is a hyperactive Jupiter family comet (JFC). Icy grains driven out by CO$_2$ sublimation have been proposed as a driver of hyperactivity, but the CO$_2$ abundance of HMP has not been measured. From our [\ion{O}{1}] line ratio measurements, we find a CO$_2$/H$_2$O ratio for HMP of $22.9\pm1.4\%$. We compare the CO$_2$/H$_2$O ratios to the active fractions of the nine comets (including HMP) in the literature that have data for both values. We find no correlation. These findings imply that CO$_2$ sublimation driving out icy grains is not the only factor influencing active fractions for cometary nuclei.

\end{abstract}


\section{Introduction} \label{sec:intro}

Comets are primitive bodies that coalesced during solar system formation and have undergone relatively little change in the past 4.5 billion years. Thus, their compositions provide essential windows into our solar system’s early formational stages. The dynamical reservoirs of comets typically are either the Oort cloud or the Scattered Disk, which inhabit cold environments far from the Sun. However, during passages through the inner solar system, comets approach the Sun and undergo heating and resultant sublimation. 

Jupiter-family comets (JFCs) are short period ($<20$ year orbits) comets that have had their orbits modified via close passages to Jupiter. As such, they likely experience a significant amount of thermal alteration. Understanding the effects of repeated perihelion passages on JFCs can yield insights into the level of thermal evolution that comets have experienced over the 4.5 billion years since formation. Hypervolatiles are one way of tracing that thermal evolution.

Hypervolatiles are chemical species with low vacuum sublimation temperatures of less than 80 K. Observations of hypervolatiles such as CO$_2$ and CO can yield constraints on the long term evolution of cometary composition due to repeated perihelion passages. \citet{AHearn2012} studied historical observations of CO$_2$ and CO for trends with dynamics, noting that CO$_2$/H$_2$O ratios appear to be depleted for comets with perihelion distances lower than 1 AU. However, they found no evidence for a strong correlation between CO abundance and perihelion distance. In addition to being a possible indicator of past thermal evolution, CO$_2$ is also an important driver of cometary activity, and may play an important role in driving hyperactive comets \citep[][and references therein]{SunshineFeaga2021}.

Hyperactive comets produce more water vapor than can be accounted for by their size.  A current leading hypothesis suggests that icy grains driven out by CO$_2$ sublimation could be a primary driver of hyperactivity \citep{AHearn2011,SunshineFeaga2021}. Specifically, the icy grain hypothesis proposes that a reservoir of volatile ice sublimates inside the nucleus of the comet. This sublimated material leaves the nucleus, bringing along with it less volatile solids. The icy grains then slowly sublimate once they are in the coma, becoming a second source of sublimation (the other being the surface of the nucleus). Typically, the initial highly volatile sublimating material is CO$_2$ and the icy grains are H$_2$O, but \citet{SunshineFeaga2021} suggest that the same process could occur for other pairs of highly volatile gasses and lower volatility ice grains. EPOXI’s flyby of comet 103P/Hartley 2 resulted in direct evidence for icy grains sublimating in the coma, and centimeter-sized ice chunks were directly observed in the inner coma \citep{AHearn2011, Kelley2013}. In addition, water ice absorption was observed in the coma and was spatially co-located with the CO$_2$ emission from the small lobe of the nucleus \citep{AHearn2011}. These observations imply that CO$_2$ was driving out the icy grains, providing more surface area for sublimation \citep{AHearn2011,SunshineFeaga2021}. Measuring the CO$_2$ abundances (also referred to as the CO$_2$/H$_2$O ratio) of hyperactive and highly active comets to determine if they have a pattern of high CO$_2$ abundances \citep[higher than the cometary median of $\sim12-17 \%$,][]{Ootsubo2012,HarringtonPinto2022} can provide evidence for or against the icy grain hypothesis. However, CO$_2$ is rarely studied compared to other volatiles.

While CO$_2$ observations are key to understanding hyperactivity in comets, CO$_2$ can only be directly observed from space due to severe telluric absorption, which limits its study \citep[e.g.][]{Combes1988, Tozzi1998, Feaga2007, AHearn2011,  Feaga2014, McKay2016, McKay2019, combi2020, HarringtonPinto2022, Gicquel2023}. Given the limited observing time available on space-based platforms, indirect observation of CO$_2$ from ground-based observatories requires a proxy.

One such proxy is the [\ion{O}{1}] line ratio, which we define as the flux ratio of the [\ion{O}{1}] 5577 Å line to the sum of the [\ion{O}{1}] 6300 Å and [\ion{O}{1}] 6364 Å lines:
\begin{equation} \label{eq:olr}
    \frac{I_{5577}}{I_{6300}+I_{6364}}
\end{equation}
where $I_{5577}$ is the flux of the 5577 Å line, $I_{6300}$ is the flux of the 6300 Å line, and $I_{6364}$ is the flux of the 6364 Å line. The [\ion{O}{1}] line ratio has been shown to be sensitive to cometary CO$_2$ abundances \citep{FestouFeldman1981, McKay2012, McKay2013, McKay_2021, Decock2013}. 

\begin{figure}[ht!]
\plotone{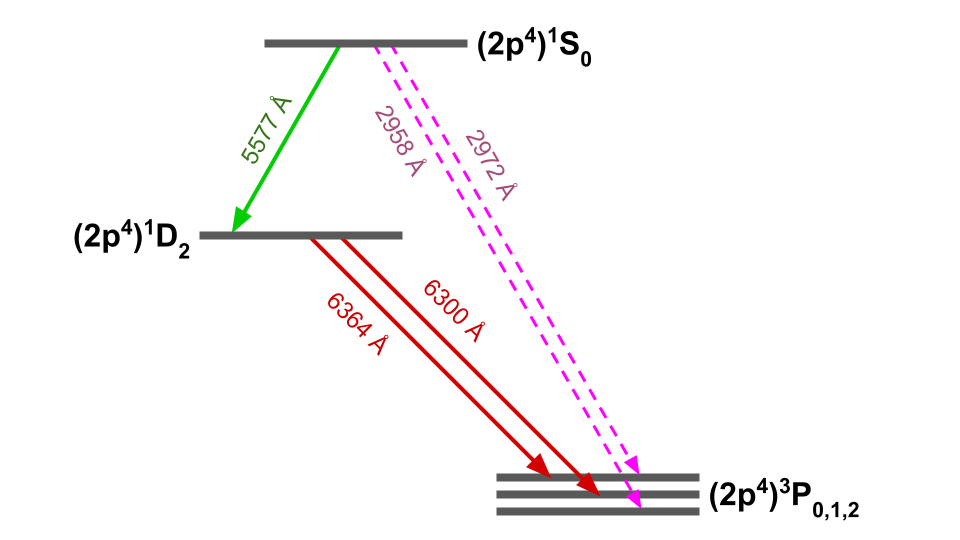}
\caption{Grotrian diagram for several forbidden oxygen lines. 90-95$\%$ of emission occurs along the solid line pathway. The other 5-10$\%$ of emission follows the dashed lines. The green solid line is the 5577 Å line, the red doublet lines are the 6300 Å and 6364 Å lines, and the ultraviolet doublet (dashed lines) are the 2972 Å and the 2958 Å lines. Adapted from \citet{Decock2013}. 
\label{fig:grot}}
\end{figure}

However, the photochemistry responsible for the release of [\ion{O}{1}] into the coma is not yet fully understood and requires more study in order for the [\ion{O}{1}] line ratio to be used as a reliable proxy for CO$_2$ production in comets \citep{Huestis2008, BhardwajRaghuram2012, McKay2013, McKay2015}.  Photodissociation of H$_2$O and CO$_2$ releases excited [\ion{O}{1}] in the $^1$S and $^1$D states, which then radiatively decay, to create emission features at 5577~\AA~($^1$S) and 6300~\AA~and 6364~\AA~($^1$D) (Fig. \ref{fig:grot}). Photodissociation of CO$_2$ releases more excited [\ion{O}{1}] in the $^1$S state relative to the $^1$D state compared to H$_2$O, causing higher [\ion{O}{1}] line ratios for cases where CO$_2$ is the dominant source of [\ion{O}{1}] in the coma~\citep{FestouFeldman1981}. Despite the uncertainty still surrounding our understanding of [\ion{O}{1}] photochemistry in cometary comae, an empirical understanding derived in association with [\ion{O}{1}] line ratio analyses has produced some agreement with CO$_2$ abundances measured using space-based observations \citep{McKay2015, McKay2016}. \citet{HarringtonPinto2022} did not find any systematic difference between comets where the CO$_2$ abundance was directly measured and those in which the CO$_2$ abundance was inferred using the oxygen line ratio.

Comet 45P/Honda-Mrkos-Pajdušáková (hereafter referred to as HMP) is a small hyperactive JFC \citep{Lamy1999, SunshineFeaga2021}. Although the exact definition of hyperactivity is debated, HMP has a much higher water active area than is measured for most comets, and thus we refer to it as hyperactive \citep{Lamy1999}. \citet{DiSanti2017} measured a CO/H$_2$O abundance ratio of (0.60$\pm$0.04)$\%$ on January 6-8, 2017, and \citet{DelloRusso2020} gave an upper limit of 3.7$\%$ on February 19, 2017. Despite additional abundances for other volatiles, HMP's CO$_2$ abundance is unknown \citep{DiSanti2017, DelloRusso2020}. 

We present an analysis of the forbidden oxygen line emission from HMP. We report the measured [\ion{O}{1}] line ratio and inferred CO$_2$/H$_2$O ratio, as well as water production rates and active fractions. Section \ref{sec:obsana} details our observations and data reduction, and section \ref{sec:results} presents our results. Section \ref{sec:comp} provides a comparison to other comets and a discussion of our results. Section \ref{sec:conclu} summarizes our conclusions.

\section{Observations and Data Analysis} \label{sec:obsana}

We obtained spectra of HMP in February 2017 using the Tull Coud\'e Spectrograph mounted on the 2.7-meter Harlan J. Smith telescope at McDonald Observatory. The instrument covers a spectral range from 3400 to 10,000~\AA, with a resolving power of $R \equiv \lambda /\Delta \lambda \approx 60,000$ \citep{Tull1995}.  The slit is 1.2\arcsec $\times$ 8\arcsec, which translates to a projected distance at the comet of approximately 140 $\times$ 950 km over the course of our observing run.

\begin{deluxetable}{cccccccc}
\tablecaption{Observing Log} \label{tab:observations}
\tablehead{\colhead{} & \colhead{} & \colhead{Geocentric} & \colhead{Geocentric} & \colhead{Heliocentric} & \colhead{Heliocentric} & \colhead{} & \colhead{}
\\ 
\colhead{Date} & \colhead{Number of} & \colhead{Distance} & \colhead{Velocity} & \colhead{Distance} & \colhead{Velocity} & \colhead{Telluric Std} & \colhead{Flux Std} \\
\colhead{} & \colhead{Exposures} & \colhead{(AU)} & \colhead{(km/s)} & \colhead{(AU)} & \colhead{(km/s)} & \colhead{} & \colhead{} } 
\startdata
Feb 21, 2017 & 5 & 0.1525 &  19.54 &  1.12 &  24.58 &  HR 4633 &  109 Vir \\
Feb 22, 2017 & 2 &  0.1640 &  20.20 &  1.13 &  24.53 &  HR 5778 &  HD 93521 \\
Feb 23, 2017 & 3 & 0.1758 &  20.77 &  1.15 &  24.47 &  HR 6033 &  Zeta Ophiuchi \\
\enddata
\end{deluxetable}

Table \ref{tab:observations} shows our observing log for the three observing nights, UT February 21-23, 2017. We used the solar port, which feeds daylight directly into the spectrograph, to obtain a solar reference spectrum. We observed quartz and ThAr lamps for flatfielding and wavelength calibration respectively. We also observed a fast rotating star for removal of telluric features and as a flux calibration standard, which are given in Table \ref{tab:observations}. Each night we obtained multiple 1800-second exposures on the comet. As shown in Table \ref{tab:observations}, we took five exposures on February 21, two exposures on February 22, and three exposures on February 23. Before extraction, we combined the comet spectra. For more detail on data reduction methods for Tull Coud\'e spectra, see \citet{CochranCochran2002}.

\begin{figure}[ht!]
\plotone{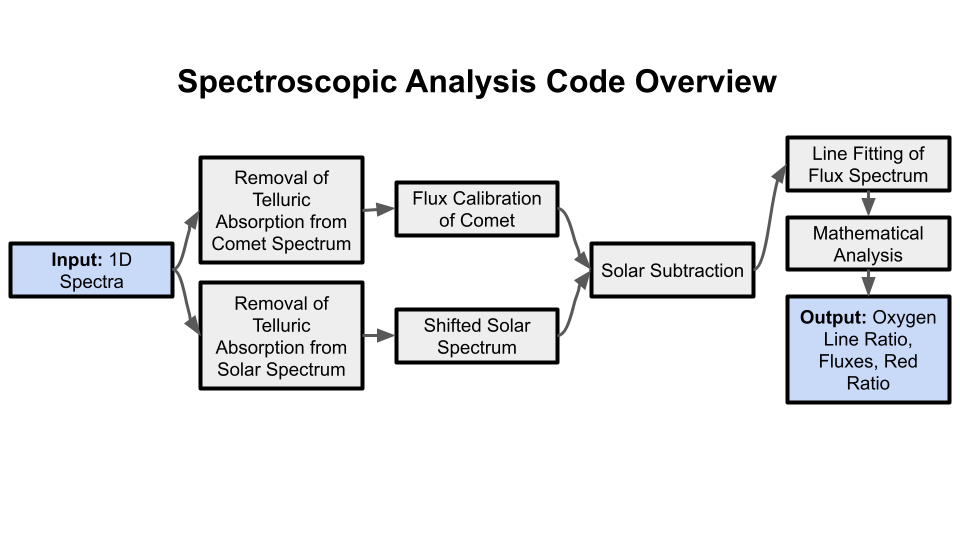}
\caption{Diagram showing the structure of the code we used to calculate the [\ion{O}{1}] line ratio from input spectra. The [\ion{O}{1}] line ratio is given by equation \ref{eq:olr}, and the red line ratio is given by equation \ref{eq:redlineratio}.
\label{fig:code}}
\end{figure}

After extraction, we performed telluric correction, flux calibration, and solar continuum removal. We fit the line profiles of emission features using a methodology similar to the process outlined in \citet{McKay2012}; however, we used a Python code that uses pyspeckit\footnote{Documentation: https://pyspeckit.readthedocs.io} instead of the previous IDL code that used mpfit\footnote{Documentation: https://www.l3harrisgeospatial.com/docs/mpfit.html}. We then integrated over the model line profiles to determine fluxes for each line, which were then used to calculate [\ion{O}{1}] line ratios. See Fig. \ref{fig:code} for an overview of our reduction and analysis algorithm. We calculated a weighted average of the [\ion{O}{1}] line ratios for each date.

We found no significant flux contamination by other chemical species. We compared our spectra to the line intensities and positions in the line atlas presented in \citet{CochranCochran2002}. The 5577 Å line is in the middle of the $\Delta \nu=-1$  C$_2$ band. There is a weak line underneath our feature, but examination of the spectrum shows no detection of the stronger lines within 10 Å, thus the line underneath is negligible ($<10\%$). There are NH$_2$ features near the red 6300 Å line. However, the closest NH$_2$ lines are separated and/or weak. The closest line (in the X-X (0, 12, 0) band) is 0.12 Å away from the 6300 Å line, which is far enough away to be completely resolved from the oxygen line at the resolving power of the Tull Coud\'e ($\Delta \lambda=0.1$ Å in this region) and is a weak line. In addition, we do not see stronger lines blueward of our 6300 Å line in the 6299 Å area, which indicates that the 6300 Å line is not contaminated by the weaker NH$_2$ X-X (0, 12, 0) band line. The 6300 Å line is also embedded in the NH2 A-X (0,3,0) band. However, the closest NH$_2$ line in this band to the 6300 Å line is 0.23 Å away and does not contaminate the 6300 Å line at our observed resolving power. The closest (unidentified) line to the 6364 Å line is 0.4 Å away, and so does not contaminate the 6364~\AA~line flux.

We used the flux ratio of the 6300 \AA~line to the 6364 \AA~line (the red line ratio) as a check on our calibration:
\begin{equation} \label{eq:redlineratio}
    2.997=\frac{I_{6300}}{I_{6364}}
\end{equation}
The red line intensity ratio should always be equal to 2.997, irrespective of coma chemistry or physics \citep{Galavis, Storey}. The intensity ratio comes from the branching ratio into the different transitions. Accordingly, it does not depend on cometary characteristics. All of our red line ratio measurements were consistent with the theoretical value.

We use the [\ion{O}{1}]6300~\AA~line flux as a proxy for water production rates. To translate from the [\ion{O}{1}]6300~\AA~line flux to a water production rate, we use a computationally simple Haser model \citep{Haser1957} which has features that emulate the vectorial formalism~\citep{Festou1981}. The Haser model approximates the coma as an isotropically expanding gas cloud and accounts for photodissociation losses. The model calculates the number density of [\ion{O}{1}] as a function of nucleocentric distance. We then use a numerical integrator (QPINT1D \footnote{Documentation: https://www.nv5geospatialsoftware.com/docs/qpint1d.html}) to convert from the spatial density function to a column density along the line of sight for each point in the field of view \citep{Morgenthaler2001,Morgenthaler2007}. Given these values, we find the production rate of water such that the Haser model's predicted flux for the [\ion{O}{1}]6300~\AA~line matches the observed flux. See \citet{McKay2012} for more details on the specific Haser model method we use.

We infer the CO$_2$ abundance from our [\ion{O}{1}] observations using the following equation:
\begin{equation} \label{eq:co2abun}
    \frac{N_{CO_2}}{N_{H_2 O}}=\frac{RW^{red}_{H_2 O}-W^{green}_{H_2 O}}{W^{green}_{CO_2}-RW^{red}_{CO_2}}
\end{equation}
where $N$ is the column density, $R$ is the [\ion{O}{1}] line ratio, and $W$ is the release rate \citep{McKay2012}. 

\begin{deluxetable}{lccccc}

\tablecaption{[\ion{O}{1}] Release Rates} \label{tab:releaserates}

\tablehead{\colhead{Parent Species} & \colhead{[\ion{O}{1}] State} & \colhead{W$^a$} & \colhead{W$^b$} & \colhead{W$^c$} & \colhead{W$^d$}} 

\startdata
H$_2$O & $^1$S & 0.64 & 0.64 & 2.6 & 2.82 \\
H$_2$O & $^1$D & 84.4    & 84.4 & 84.4 & 53.8\\ 
CO$_2$ & $^1$S & 50.0 & 33.0 & 72.0 & \\
CO$_2$ & $^1$D & 75.0 & 49.5 & 120.0 & \\
CO & $^1$S & 4.0 & 4.0 & 4.0 &  \\
CO & $^1$D & 5.1 & 5.1 & 5.1 &   \\
\enddata

\tablecomments{$^1$S is the green line, and $^1$D is the red line. Table reproduced from \citet{McKay2016}.\\ $a$ Release rates in 10$^{-8}$ s$^{-1}$ from \citet{McKay2015} set A.\\ $b$ Release rates in 10$^{-8}$ s$^{-1}$ from \citet{McKay2015} set B.\\ $c$ Release rates in 10$^{-8}$ s$^{-1}$ from \citet{BhardwajRaghuram2012}. \\ $d$ Release rates in 10$^{-8}$ s$^{-1}$ from \citet{kawakita2022}, using the different branching ratios case. \\ We include the last two sets for comparison to our adopted release rates.}

\end{deluxetable}

While not well constrained, reasonable empirical values for the release rates exist in the literature. Four sets of these release rates are given in Table \ref{tab:releaserates}. \citet{McKay2015} used observations of C/2009 P1 (Garradd) to derive sets A and B. Set C is from \citet{BhardwajRaghuram2012}. Set D is from \citet{kawakita2022}, using the different branching ratio case. Release rates B tend to reproduce space-based CO$_2$/H$_2$O ratios most accurately. We present results using both sets A and B to show the sensitivity of our results to the adopted release rates.

We include the effects of CO and collisional quenching of the $^1$D state in our calculations for both sets of release rates. To include CO, we used the more detailed form of equation \ref{eq:co2abun} given as equation 6 in \cite{McKay2012}. Inclusion of CO lowers the CO$_2$ abundance inferred using Eq. \ref{eq:co2abun} \citep{McKay2012}. In general, the contribution of [\ion{O}{1}] from CO is small unless the CO/CO$_2$ ratio is much greater than unity, as shown by~\cite{McKay2019} and~\cite{Raghuram2020}. 
The CO/H$_2$O ratio for HMP is 0.6 $\pm$ 0.04$\%$ \citep{DiSanti2017}, and we find a negligible effect for CO on our inferred CO$_2$ abundances. 

Our inferred CO$_2$ abundance from the [\ion{O}{1}] line ratio does not account for molecular oxygen as a parent molecule for [\ion{O}{1}]. However, \cite{McKay2018} showed that oxygen line fluxes would only be affected for very large O$_2$ abundances, so it is unlikely including O$_2$ would significantly affect our results. There is no reported measurement of molecular oxygen for HMP, nor for most comets.  We calculated collisional quenching effects on our measured [\ion{O}{1}] line fluxes as outlined in \citet{McKay2015} but found it had a negligible effect on the [\ion{O}{1}] line ratio.

Our reported uncertainties reflect 1$\sigma$ stochastic errors dominated by Poisson statistics. We account for additional systematic uncertainties due to the flux calibration processes as outlined in~\cite{McKay2014}. 
These systematic uncertainties only influence our water production rates and not the [\ion{O}{1}] line ratio measurements, as only the water production rates are sensitive to the absolute flux calibration.

\section{Results} \label{sec:results}

\begin{figure}[ht!]
\plotone{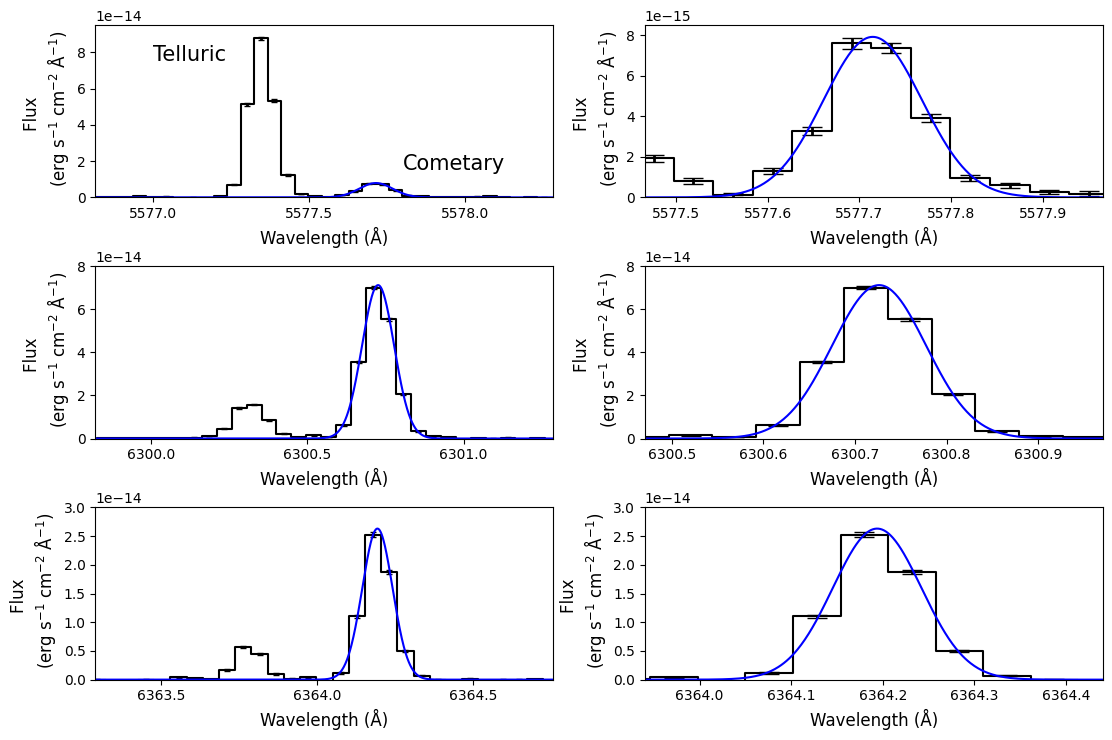}
\caption{Plots of the fitted Gaussian line profiles (blue lines) for the spectra obtained on UT February 21. The left column shows the cometary line (redward) and the telluric line (blueward). The right column is zoomed-in on just the fitted cometary line. The first row is the 5577 \AA~line, the second row is the 6300 \AA~line, and the third row is the 6364 \AA~line. 
\label{fig:bigfig}}
\end{figure}

Fig. \ref{fig:bigfig} shows an example of the spectra in the regions containing the [\ion{O}{1}] emission lines, with best fit line profiles overplotted. The left column shows the fitted cometary line in context of a wider spectral range, with the unfitted telluric line blueward of the cometary line (the separation between the telluric and cometary lines was large enough that there was no blending of the line profiles---see Table~\ref{tab:observations}). The right column presents a zoomed-in view, showing only the cometary line. The first row is the 5577 \AA~line, the second row is the 6300 \AA~line, and the third row is the 6364 \AA~line.

\begin{deluxetable}{ccc}

\tablecaption{Results for HMP by Date} \label{tab:datedata}

\tablehead{\colhead{Date} & \colhead{[\ion{O}{1}] Line} & \colhead{Water Production Rate} \\ 
\colhead{} & \colhead{Ratio} & \colhead{($\times 10^{27}$ mol s$^{-1}$)}
} 

\startdata
Feb 21, 2017 & 0.0679$\pm$0.0042 & 1.90 $\pm$ 0.02 \\
Feb 22, 2017 & 0.090$\pm$0.013 & 2.64 $\pm$ 0.04 \\
Feb 23, 2017 & 0.074$\pm$0.019 & 1.88 $\pm$ 0.02 \\ 
\enddata

\end{deluxetable}

We calculated a weighted mean [\ion{O}{1}] line ratio for HMP from all three dates of $0.077 \pm 0.004$. Values for individual dates are given in Table \ref{tab:datedata}.  Using release rates B, this [\ion{O}{1}] ratio yields a CO$_2$ to H$_2$O ratio of $22.9\pm1.4\%$. Using release rates A, this [\ion{O}{1}] ratio yields a CO$_2$ to H$_2$O ratio of $15.2\pm0.9\%$. We prefer release rates B, since those rates reproduce space-based observations of CO$_2$ for C/2012 K1 (PanSTARRS) better \citep{McKay2016}. We calculate a value of $\sim6 \%$ using release rates set C from \citet{BhardwajRaghuram2012}. We do not use this value, as that set of release rates underestimated the CO$_2$ abundances of 186P/Garradd and C/2012 K1 (PanSTARRS) \citep{McKay2015, McKay2016}. Using the water release rates set D from \citet{kawakita2022}, depending on which CO$_2$ release rates are used, we obtain a CO$_2$ abundance of $1-8\%$. We include these values for comparison to our preferred value of $22.9\pm1.4\%$.

We determined water production rates using a Haser model to reproduce the 6300 \AA~line integrated flux. Using the values for each date (given in Table \ref{tab:datedata}), we found an average water production rate of $(2.14 \pm 0.16) \times 10^{27}$ mol s$^{-1}$.  Using the slow rotator approximation and the sublimation model of \citet{CowanAHearn1979}\footnote{Small Bodies Node model: https://ice-sublimation-tool.astro.umd.edu/}, we calculated a water vaporization rate of $2.25 \times 10^{17}$ mol cm$^{-2}$ s$^{-1}$ at the time of our observations. We use the slow rotator model because cometary nuclei have low thermal inertia \citep{Groussin2013, Bodewits2014,McKay2018}. Using our measured production rate and calculated vaporization rate, we calculated an active area of $0.95 \pm 0.07$ km$^2$. Using the radius value of 0.65$\pm$0.05 km from radar observations described in \citet{LejolyHowell2017} to compare the active area to the entire surface area of the comet, we calculate a water active fraction of $17.9 \pm 3.1 \%$. We assumed Poisson distributed statistics for our data and propagated the values by using uncertainties given in the cited papers as necessary to produce the uncertainties on our stated values.

\section{Discussion} \label{sec:comp}

\subsection{Comparison to other HMP observations}
\citet{combi2020} measured a water production rate of $\sim 3\times10^{27}$ mol s$^{-1}$ for the 2017 apparition at heliocentric distances $R_h$=1.0 to 1.1 AU. This is $\sim 50\%$ higher than our value of $(2.14\pm0.16)\times10^{27}$ mol s$^{-1}$. \citet{combi2020} determined water production rates using SoHO SWAN observations of the Lyman alpha line, which sample a larger portion of the coma compared to our observations. The fact that our observed water production rate is slightly lower than that of \citet{combi2020} could support the icy grain hypothesis since the icy grains are lifted off of the nucleus and begin to sublimate while simultaneously expanding outwards. Because our observations only cover the inner coma, we may be missing the water sublimation from icy grains in the outer portions of the coma sampled by \citet{combi2020}. If the icy grain hypothesis is valid, then a larger field of view would result in higher observed water production rates, as discussed in \cite{McKay2015}. This could explain \citet{combi2020}'s water production rate being higher than ours.

\citet{DelloRusso2020} studied water directly in the infrared. They had a similarly narrow slit to our observations and thus should have a similar water production rate to our value based on the above discussion. They measured a water production rate of $\sim 2\times10^{27}$ at $R_h=1.10$ AU, which is consistent with our value of $(2.14\pm0.16)\times10^{27}$ mol s$^{-1}$. \citet{DiSanti2017} also observed in the infrared and found a much higher water production rate of 2-3$\times 10^{28}$ mol s$^{-1}$ at $R_h=0.55$ AU. However, their observations were obtained much closer to perihelion. Since HMP was closer to the Sun at the time of those observations, the insolation was much higher. Therefore, they are not as directly comparable to our observation as \citet{combi2020} and \citet{DelloRusso2020} are.

\subsection{Implications for Icy Grain Hypothesis}
\subsubsection{Comparison to 46P/Wirtanen and 103P/Hartley 2}

\begin{deluxetable}{ccc}

\tablecaption{Comparison of HMP to Wirtanen and Hartley 2} \label{tab:wirthart}

\tablehead{\colhead{Comet Name} & \colhead{CO$_2$/H$_2$O ($\%$)} & \colhead{CO/H$_2$O ($\%$)}
} 

\startdata
45P/HMP & 22.9±1.4 & 0.6±0.04$^a$ \\
46P/Wirtanen & 14.7±1.6$^b$ & $<$0.54$^b$ \\
103P/Hartley 2 & 10-20$^c$ & 0.15-0.45$^d$ \\ 
\enddata

\tablecomments{\\$a$ \citet{DiSanti2017} \\
$b$ \citet{McKay_2021}\\ 
$c$ \citet{AHearn2011} \\ 
$d$ \citet{Weaver2011} }

\end{deluxetable}

Table \ref{tab:wirthart} compares our CO$_2$ abundance and \citet{DiSanti2017}'s CO abundance for HMP to the values for 46P/Wirtanen (hereafter Wirtanen) and 103P/Hartley 2 (hereafter Hartley 2), two other hyperactive comets~\citep{AHearn2011,Lamy1999,Groussin2004}. These two comets are small hyperactive JFCs that are similar in composition, size, and activity drivers. 
HMP has a high CO$_2$ abundance similar to both Wirtanen and Hartley 2. These comets have high active fractions and high CO$_2$ abundances. This connection between high active fraction and high CO$_2$ would support the icy grain hypothesis with CO$_2$ acting as a primary driver of hyperactivity. In addition, the high CO$_2$ abundances in these and other JFCs imply a minimal effect of perihelion passages on their CO$_2$ content.

It has been noted that there may be a correlation between hyperactivity and the D/H ratio in water ~\citep{Lis2019}. This correlation is supported by these three hyperactive comets, all of which have low D/H ratios consistent with VSMOW \citep[Vienna Standard Mean Ocean Water,][]{Hartogh2011,Lis2013,Lis2019}. 

It has been hypothesized that Wirtanen and Hartley 2 may be members of a new JFC compositional subfamily \citep{McKay_2021}. HMP seems to fit into this potential subfamily. In addition to these three comets being hyperactive and having high CO$_2$ abundances as noted above, they also all have low CO abundances, in addition to similar abundances of other trace volatiles (CO, CH$_3$OH, and C$_2$H$_6$) ~\citep{McKay_2021, Khan2023}. Could this possible family be reflective of an ancient breakup of a larger object, or is it indicative of composition in a specific section of the protoplanetary disk? If a large number of comets fall into this subfamily, the latter becomes more likely.

Taken together, the high CO$_2$ abundances and active fractions of these comets seem to support CO$_2$ sublimation driving out icy grains. However, we should broaden our comparison to a larger number of comets.

\subsubsection{Comparison to Other Comets}

\begin{deluxetable}{lrrrrrr}

\tablecaption{Observed Parameters Related to Cometary Active Fractions} \label{tab:bigtable1}

\tablehead{
\colhead{Comet} &\colhead{Radius} &\colhead{Surface} &\colhead{Heliocentric} &\colhead{Water Production} &\colhead{CO$_2$ Production} &\colhead{CO$_2$} \\
\colhead{Name} &\colhead{(km)} &\colhead{Area (km$^{2}$)} &\colhead{Distance (AU)} &\colhead{Rate (mol s$^{-1}$)} &\colhead{Rate (mol s$^{-1}$)} &\colhead{Abundance ($\%$)}
} 

\startdata
1P &	 &	 &	 0.79 $^k$ &	 &	 &	$2.5\pm0.5$ $^i$ \\
\hline
22P &	$1.59^{+0.46}_{-0.36}$ $^b$ &	$32\pm18$ &	1.61 $^e$ &	$(1.93\pm0.19)\times10^{27}$ $^e$ &	$(3.89\pm0.39)\times10^{26}$ $^e$ &	$20.1\pm2.9$  \\
&	&	&	1.61 $^e$ &	$(5.94\pm0.6)\times10^{27}$ $^e$ &	$(1.10\pm0.11)\times10^{27}$ $^e$ &	$18.5\pm2.6$  \\
&	&	&	2.42 $^e$ &	$(1.22\pm0.13)\times10^{27}$ $^e$ &	$(1.34\pm0.14)\times10^{26}$ $^e$ &	$11.0\pm1.6$  \\
&	&	&	2.43 $^e$ &	$1.05\pm0.11)\times10^{27}$ $^e$ &	$(7.40\pm0.80)\times10^{25}$ $^e$ &	$7.1\pm1.1$  \\
&	&	&	2.43 $^e$ &	$(1.70\pm0.17)\times10^{27}$ $^e$ &	$(7.00\pm0.80)\times10^{25}$ $^e$ &	$4.1\pm0.6$  \\
\hline
45P$^a$ &	$0.65\pm0.05$ $^c$ &	$5.31\pm0.82$ &	1.13 &	$(2.14\pm0.16)\times10^{27}$ &	&	$22.9\pm11.4$  \\
\hline
46P &	&	&	1.13 $^h$ &	$(4.61^{+1.52}_{-0.65})\times10^{27}$ $^h$ &	&	$14.7\pm1.6$ $^h$  \\
\hline
67P &	&	&	1.84 $^e$ &	$(6.20\pm0.63)\times10^{26}$ $^{e, \beta}$ &	$(4.30\pm0.50)\times10^{25}$ $^e$ &	$6.9\pm1.1$  \\
&	&	&	1.75 $^f$ &	$(4.91\pm0.57)\times10^{26}$ $^{f, \gamma}$ &	$(1.45\pm2.82)\times10^{25}$ $^f$ &	$1.1\pm2.2$  \\
&	&	&	1.24  $^f$ &	$(3.96\pm0.70)\times10^{27}$ $^{f,\delta}$ &	$(1.41\pm0.74)\times10^{27}$ $^f$ &	$7.9\pm4.8$  \\
&	&	&	1.75  $^f$ &	$(9.78\pm5.35)\times10^{25}$ $^{f,\epsilon}$ &	$(2.49\pm1.54)\times10^{26}$ $^f$ &	$10.4\pm7.4$  \\
\hline
103P &	$0.57\pm0.08$ $^d$ &	$4.08\pm1.15$ &	1.24  $^d$ &	$(1.1\pm0.1)\times10^{28}$ $^d$ &	&	$15\pm5$ $^j$ \\
\hline
118P &	$3.29^{+2.71}_{-1.53}$ $^b$ &	$136\pm224$ &	2.21 &	$1.05\times10^{27}$ $^{g, \alpha}$ &	$(2.57\pm0.20)\times10^{26}$ $^g$ &	$24.5\pm1.9$ \\
&	&	&	2.01 &	$1.39\times10^{27}$ $^{g, \alpha}$ &	$(3.31\pm0.20)\times10^{26}$ $^g$ &	$23.9\pm1.4$ \\
&	&	&	2.18 $^e$ &	$(1.43\pm0.15)\times10^{27}$ $^e$ &	$(2.18\pm0.22)\times10^{26}$ $^e$ &	$15.3\pm2.2$ \\
&	&	&	2.18 $^e$ &	$(9.97\pm0.10)\times10^{26}$ $^e$ &	$(2.99\pm0.30)\times10^{26}$ $^e$ &	$30.0\pm4.3$ \\
\hline
144P &	$0.81\pm0.04$ $^b$ &	$8.2\pm0.8$ &	1.70 $^e$ &	$(3.76\pm0.38)\times10^{27}$ $^e$ &	$(5.57\pm0.56)\times10^{27}$ $^e$ &	$14.8\pm2.1$  \\
&	&	&	1.70 $^e$ &	$(3.33\pm0.33)\times10^{27}$ $^e$ &	$(5.15\pm0.52)\times10^{26}$ $^e$ &	$15.5\pm2.2$ \\
\enddata

\tablecomments{\\
$a$ This work, $b$ \citet{Fernandez2013}, 
$c$ \citet{DiSanti2017}, 
$d$ \citet{Meech2011}, 
$e$ \citet{Ootsubo2012}, 
$f$ \citet{COMBI2020b}, 
$g$ \citet{Reach2013}, 
$h$ \citet{McKay_2021}, 
$i$ \citet{Combes1988},
$j$ \citet{Weaver2011}, and
$k$ \citet{Combes1988}.\\
$\alpha$ Water production rate derived from OH production rate.	\\
$\beta$	Remote observation.\\
$\gamma$ In situ observation, 1.75 AU pre perihelion.\\
$\delta$ In situ observation at perihelion.\\
$\epsilon$ In situ observation, 1.75 AU post perihelion.\\
The comets listed in this table are: 1P/Halley, 22P/Kopff, 45P/Honda–Mrkos–Pajdušáková, 46P/Wirtanen, 67P/Churyumov–Gerasimenko, 103P/Hartley 2, 118P/Shoemaker–Levy, and 144P/Kushida.
}

\end{deluxetable}

\begin{deluxetable}{lrrr}

\tablecaption{Calculated Values Related to Cometary Active Fractions} \label{tab:bigtable2}

\tablehead{\colhead{Comet} &\colhead{Water Vaporization} &\colhead{Water Active} &\colhead{Active} \\
\colhead{Name} &\colhead{Rate (mol cm$^{-2}$ s$^{-1}$)} &\colhead{Area (km$^{2}$)} &\colhead{Fraction ($\%$)}
} 

\startdata
1P &	 &	 &	$27^{+13}_{-7}$ $^b$ \\
\hline
22P &	$7.68\times10^{16}$ &	$2.52\pm0.25$ &	$7.9\pm4.6$ \\
&	$7.68\times10^{16}$ &	$7.74\pm0.78$ &	$24.4\pm14.3$ \\
&	$9.56\times10^{15}$ &	$12.71\pm1.31$ &	$40.0\pm23.5$ \\
 &	$9.25\times10^{15}$ &	$11.32\pm1.19$ &	$36\pm21$ \\
&	$9.25\times10^{15}$ &	$18.36\pm1.86$ &	$57.8\pm33.9$ \\
\hline
45P$^a$ &	$2.14\times10^{27}$ &	$0.95\pm0.07$ &	$17.9\pm3.1$ \\
\hline
46P &	$4.61\times10^{27}$ &	$1.41\pm0.47$ &	$36.5^{+15.5}_{-13.5}$ $^c$ \\
\hline
67P &	&	&	$3.75\pm0.44$ \\
&	$2.35\times10^{16}$ $^{\alpha}$ &	$5.49\pm2.12$ &	$15.1\pm19.9$ \\
&	$1.11\times10^{17}$ $^{\beta}$&	$15.9\pm4.6$ &	$43.9\pm56.6$ \\
&	$5.09\times10^{15}$ $^{\gamma}$&	$46.9\pm16.6$ &	$129\pm169$ \\
\hline
103P &	$1.10\times10^{28}$ $^{\delta}$&	$6.26\pm0.57$ &	$153\pm45$ \\
\hline
118P &	$1.79\times10^{16}$ &	$5.89$ &	$4.3\pm7.1$ \\
&	$3.02\times10^{16}$ &	$4.60$ &	$3.4\pm5.6$ \\
&	$1.94\times10^{16}$ &	$7.35\pm0.75$ &	$5.4\pm8.9$ \\
&	$1.94\times10^{16}$ &	$5.14\pm0.53$ &	$3.8\pm6.2$ \\
\hline
144P &	$6.27\times10^{16}$ &	$5.99\pm0.60$ &	$72.6\pm10.2$ \\
	&$6.27\times10^{16}$ &	$5.31\pm0.53$ &	$64.4\pm9.1$ 
\enddata

\tablecomments{\\$a$ This work,
$b$ \citet{JULIAN2000}, and
$c$ \citet{McKay_2021}. \\
$\alpha$	Remote observation.\\
$\beta$ In situ observation, 1.75 AU pre perihelion.\\
$\gamma$ In situ observation at perihelion.\\
$\delta$ In situ observation, 1.75 AU post perihelion.
}

\end{deluxetable}

If the icy grain hypothesis is accurate, then we would expect a statistically significant positive correlation between CO$_2$ abundance and water active fraction. We scoured the literature for data that allowed us to calculate the CO$_2$ abundance and the water active fraction which requires both a measured water production rate and size measurement. We found such data for eight comets at various heliocentric distances. Including our calculated values for HMP, the total number of comets comes to nine. These nine comets are the only ones in the literature that have values we can use to calculate both the active fraction and CO$_2$ abundance. See Tables \ref{tab:bigtable1} and \ref{tab:bigtable2} for the values used in our calculations and associated references. From the comet's effective radius, we calculated its surface area, assuming a spherical body. We used the slow rotator approximation to calculate the water vaporization rate at the heliocentric distance of each comet at the time of observation. For comets without values in the literature for water production, we used equation 1 from \citet{CochranSchleicher1993} to calculate the water production rate from OH production and heliocentric distance. Using the water production and vaporization rates, we calculated a water active fraction. Then, we used the surface area of the comet to convert from water active fraction to water active area. We then used the CO$_2$ and water production rates from the literature to calculate the CO$_2$ abundance (equivalent to the CO$_2$/H$_2$O ratio) instead of inferring it from the [\ion{O}{1}] line ratio, as we did for HMP.

Wirtanen was the only other comet whose CO$_2$ abundance was inferred from the [\ion{O}{1}] line ratio. Analysis of all other comets used directly measured H$_2$O and CO$_2$ production rates. The same release rates were used for Wirtanen as were used for HMP, thus the release rates are consistent with each other.

\begin{figure}[ht!]
\plotone{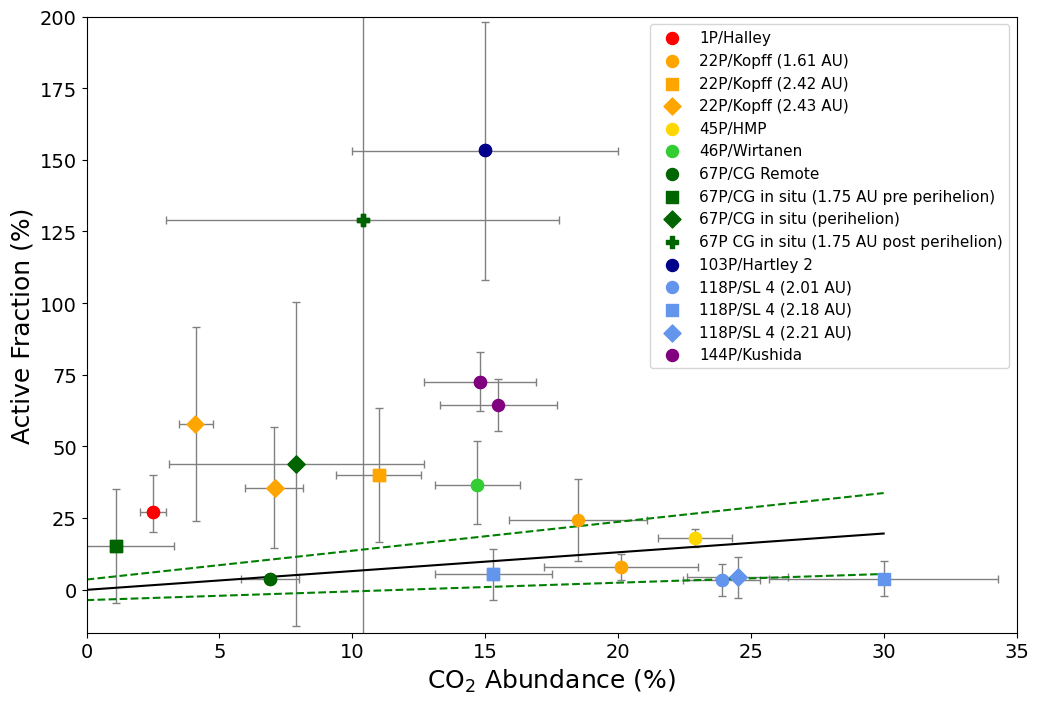}
\caption{Plot showing water active fraction vs CO$_2$ abundance. Each color is a different comet. Each marker shape within the colors is a different heliocentric distance or relation to perihelion. The solid black line is a line of best fit that was calculated via orthogonal distance regression. It has an equation of y=(0.66$\pm$0.35)x+(-0.1$\pm$3.59). The green dashed lines are $1\sigma$ bounds on the best fit line.  There is no significant correlation between active fraction and CO$_2$ abundance for this sample of comets. 
\label{fig:cometsplot}}
\end{figure}

We plotted these data in Fig. \ref{fig:cometsplot}. The CO$_2$/H$_2$O ratio is on the horizontal axis, and the water active fraction is on the vertical axis. Each color represents a different comet, and the different shapes within those colors are various heliocentric distances. The solid black line is a line of best fit calculated using orthogonal distance regression (ODR), accounting for both horizontal and vertical error bars. We treat the observations at the same heliocentric distance as different data points for the purposes of fitting. If we down weighted the similar observations, it would make the already weak positive slope even weaker. A typical method for regression analysis is the least squares approximation method. This method takes the residuals, which are the difference between a data point and the fitted value, squares each one, then sums them. It then minimizes that sum of the squared residuals to find a curve of best fit. However, typical fitting methods like least squares only account for uncertainty in the dependent variable. In our case, we have uncertainty on both the independent and dependent variables. Instead of minimizing error in just the vertical direction, ODR minimizes error perpendicular to the curve of best fit. ODR minimizes the sum of the squared orthogonal residuals.

The line of best fit has a slope of y=(0.66±0.35)x+(-0.1±3.59), though it is also clear that this linear fit does not describe the data well. The green dashed lines represent $1\sigma$ uncertainties on the line of best fit. The slope is only positive within $1\sigma$. Within $2\sigma$, we begin to encompass zero or negative slope values. This implies that there is, at best, a weak correlation between CO$_2$ abundance and water active fraction.  

As an additional test, we calculated a chi square value, using the null hypothesis that our data and the model are pulled from the same distribution. For each CO$_2$ abundance, we generated an active area using the line of best fit. We then saw if the generated active area points were statistically likely to be from the same distribution as the actual active area data. This chi square test only takes into account the vertical error bars. We found a chi square value of 128.9 with 17 degrees of freedom, resulting in a reduced chi square value of 7.58. This translates to a p-value of 3.1$\times 10^{-19}$. Our p-value is very small, which means that the probability of getting our chi square value is very low if the null hypothesis is accurate. Our p-value is much less than 0.05, thus we reject our null hypothesis. Our linear model does not predict our active fraction data well using the CO$_2$ abundance. 

We calculated another chi square value with the same null hypothesis but this time taking into account only our horizontal error bars. For each active area, we generated a corresponding CO$_2$ abundance using the line of best fit. We then saw if the generated CO$_2$ abundances were statistically likely to be from the same distribution as the actual CO$_2$ abundance data. Our reduced chi square value was 2007, which results in an even smaller p-value. We again reject the null hypothesis and determine that our linear model does not predict our CO$_2$ abundance data well using the active fraction data.

Our low p-values argue that there is not a linear relationship nor a strong correlation between CO$_2$ abundance and active fraction.  We calculated a Pearson's correlation coefficient of -0.23 without taking into account any error bars. This value implies a weak negative correlation.

The relationship may not be linear. There appears to be a peak in the active fraction data at $10-15\%$ CO$_2$ abundance. Overall, however, there seems to be no correlation. Any speculation as to a functional form without more data that demonstrate any statistical correlation is not warranted at this time. 

Taking all of our statistical calculations together, we argue that there is no correlation between active fraction and CO$_2$ abundance. The icy grain hypothesis that ice-rich grains driven out by CO$_2$ sublimating in the coma leads to hyperactivity is not supported by the results presented here. These results do not support the conclusion that CO$_2$ sublimation is solely responsible for driving out the icy grains and imply there are more factors that affect hyperactivity than just a high CO$_2$ abundance. 

\subsubsection{Effect of Possible Extended Sources of Water Production}

\begin{figure}[ht!]
\plotone{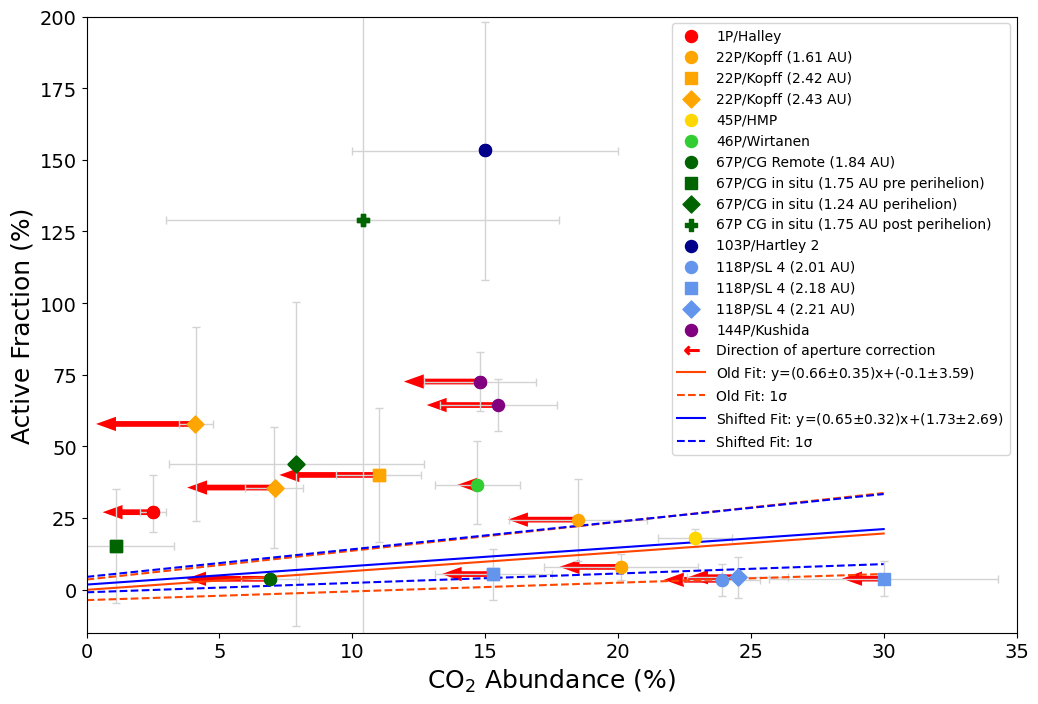}
\caption{Aperture size effects are minor, as shown in this plot of water active fraction vs CO$_2$ abundance. Each color is a different comet. Each marker shape within the colors is a different heliocentric distance or relation to perihelion. The solid orange line is a line of best fit that was calculated via orthogonal distance regression. It has an equation of y=(0.66$\pm$0.35)x+(-0.1$\pm$3.59). The orange dashed lines are $1\sigma$ bounds on the best fit line. The red arrows indicate direction and relative magnitude of the shift in CO$_2$ abundance after accounting for aperture size. The solid blue line is a line of best fit using the shifted points and was calculated using orthogonal distance regression. It has an equation of y=(0.65$\pm$0.32)x+(1.73$\pm$2.69). The blue dashed lines are $1\sigma$ bounds on the shifted best fit line.
\label{fig:shiftedplot}}
\end{figure}

If a comet has an extended source of water production, a larger aperture size will observe more of the coma, which will result in a higher water production rate and thus a lower CO$_2$ abundance relative to water. The aperture size was not accounted for in Fig. \ref{fig:cometsplot}. In Fig. \ref{fig:shiftedplot}, the red arrows give the direction and relative amount of change if we were to account for the aperture size and a potential extended source for each of these comets. We lack an understanding of how extended sources function in general, and whether these specific comets have extended sources, thus it is difficult to determine the absolute effects of aperture size on CO$_2$ abundance. Therefore the arrows serve merely as a guide for showing the qualitative effect of a hypothetical extended source in these objects. After accounting for aperture size, the CO$_2$ abundances drop, since the other instruments used had larger apertures than ours. Fitting to the shifted points, the fit does not significantly change. Overall, there is little change in slope, thus the effect of potential extended sources of water production on our conclusions is minor, and our conclusion of a lack of correlation between CO$_2$ abundance and active fraction remains unchanged.

\subsubsection{Effect of Heliocentric Distance}

The activity of comets can change with heliocentric distance. In this section we check for heliocentric distance bias in CO$_2$ abundance and water active fraction. We find no such bias for this sample of comets.

\begin{figure}[ht!]
\plotone{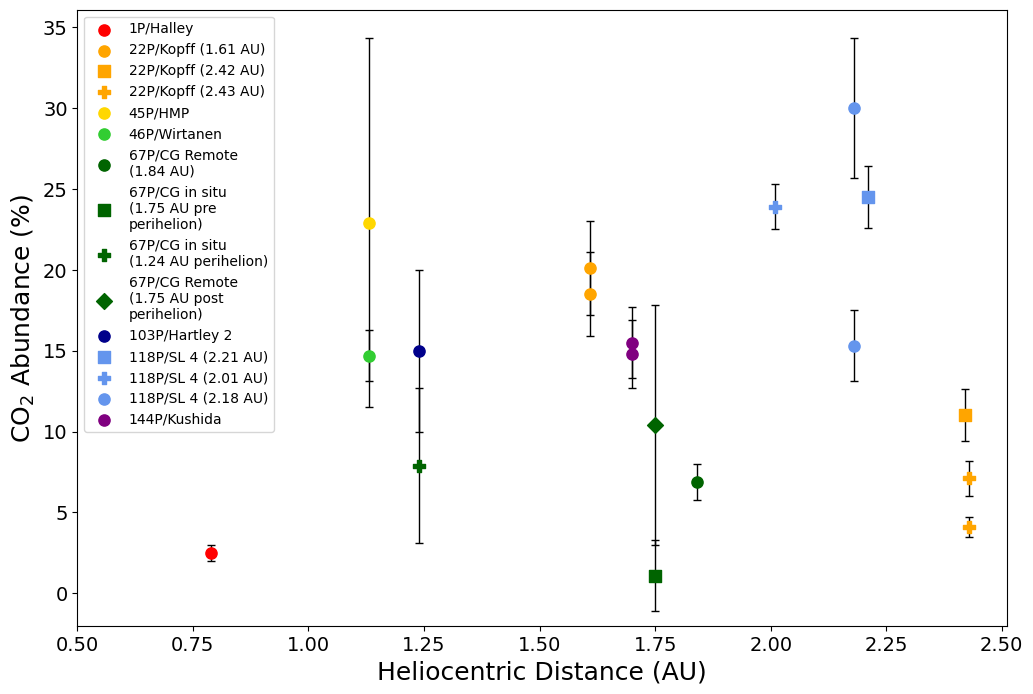}
\caption{Plot showing CO$_2$ abundance vs. heliocentric distance for our sample of comets. Each color is a different comet. Each marker shape within the colors is a different heliocentric distance or relation to perihelion. There is no correlation between CO$_2$ abundance and heliocentric distance for this sample of comets.
\label{fig:co2vhelio}}
\end{figure}

\begin{figure}[ht!]
\plotone{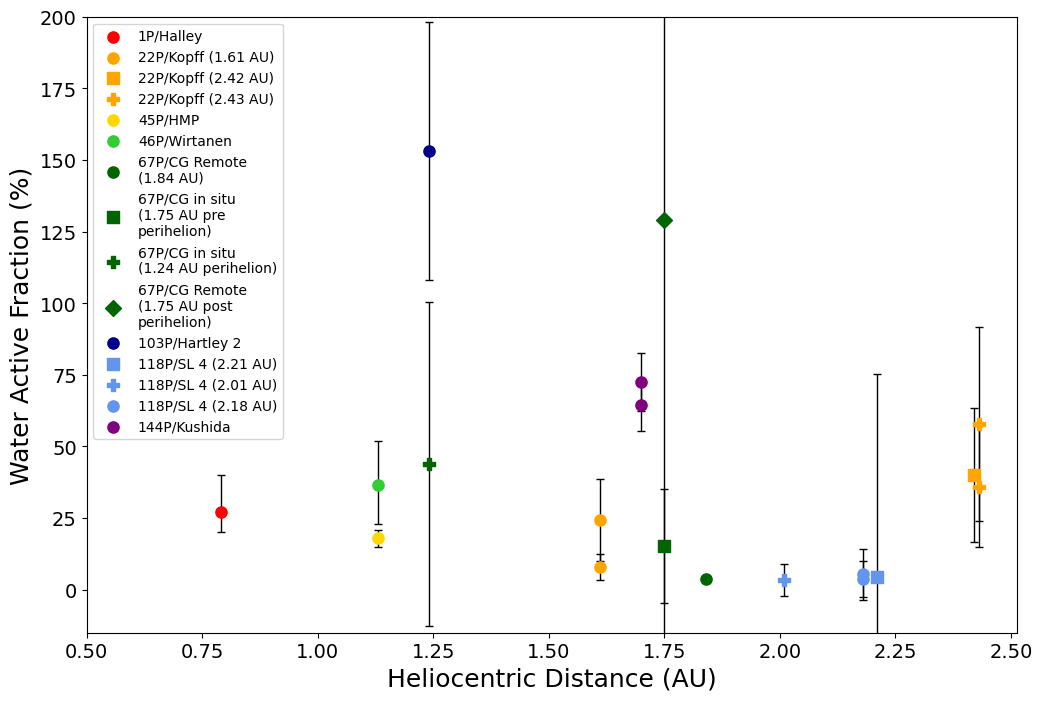}
\caption{Plot showing heliocentric distance vs. active fraction for each of the comets in our sample. Each color is a different comet. Each marker shape within the colors is a different heliocentric distance or relation to perihelion. There is no correlation between water active fraction and heliocentric distance for this sample of comets.
\label{fig:afvshelio}}
\end{figure}

\citet{HarringtonPinto2022} suggest that there is a positive power law correlation between CO$_2$ abundance and heliocentric distance. We perform a similar analysis in Fig. \ref{fig:co2vhelio}. We find that there is no correlation between CO$_2$ abundance and heliocentric distance for this sample of comets. Thus, there is no CO$_2$ abundance bias with heliocentric distance for this heliocentric distance range.  Fig. \ref{fig:afvshelio} demonstrates there is no correlation between water active fraction and heliocentric distance for this sample of comets either.

Since there is no correlation for our comet sample between either CO$_2$ abundance or water active fraction with heliocentric distance, the effects of heliocentric distance are negligible and do not affect our conclusions about the overall lack of correlation between water active fraction and CO$_2$ abundance. 

\subsubsection{22P/Kopff}
Since we have multiple points for some comets in Fig. \ref{fig:cometsplot}, we can track the active fraction compared to the CO$_2$ abundance for individual comets. 144P/Kushida has only two points that are closely spaced in both CO$_2$ abundance and active fraction, so fitting a line to those two points would not provide particularly useful constraints. In addition, the CO$_2$ abundance and active fraction remain basically constant within uncertainty, so it would be difficult to derive any useful relationship between the two. 118P/Shoemaker-Levy's data points are nearly flat in active fraction, exhibiting little variation. The data for 67P/CG has large error bars. However, 22P/Kopff has sufficient data to yield potential insights. 

\begin{figure}[ht!]
\plotone{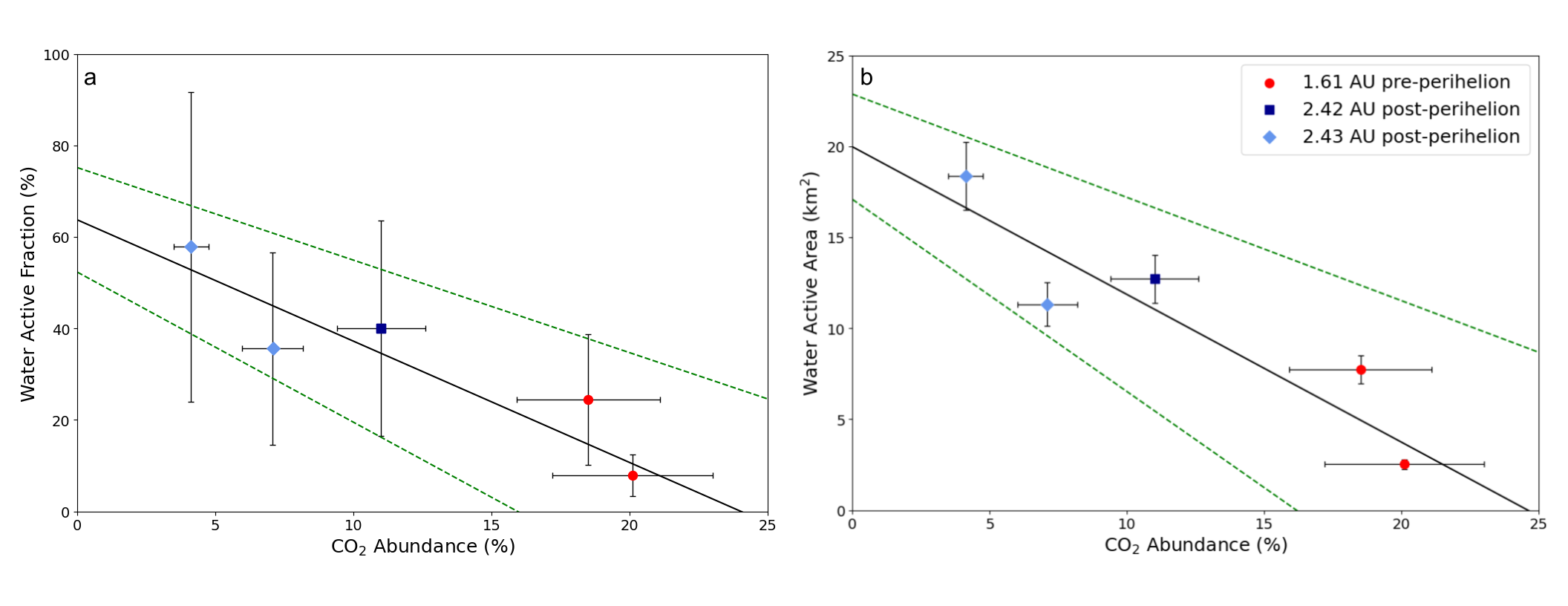}
\caption{Panel A: Plot of water active fraction vs CO$_2$ abundance for 22P/Kopff. The different marker shapes are different heliocentric distances. The rightmost two points are pre-perihelion, and the three leftmost points are post-perihelion. The solid black line is a line of best fit that was calculated via orthogonal distance regression. It has an equation of y=(-2.65$\pm$0.63)x+(63.71$\pm$11.42). The green dashed lines are $1\sigma$ bounds on the best fit line. Panel B: Plot of water active area vs CO$_2$ abundance for 22P/Kopff. The solid black line is the line of best fit with an equation of y=(-0.81$\pm$0.24)x+(19.97$\pm$2.89). The dashed green line is $1\sigma$ bounds on the line of best fit.  There are weak negative correlations for 22P/Kopff.
\label{fig:kopffplot}}
\end{figure}

Fig. \ref{fig:kopffplot} presents the water active fractions (Panel A) and active areas (Panel B) over a range of CO$_2$ abundances for 22P/Kopff both before perihelion (circles) and after perihelion (squares and diamonds). The black solid line for the active fraction has an equation of y=(-2.65±0.63)x+(63.71±11.42) and was fitted using orthogonal distance regression, which takes into account both the horizontal and vertical error bars. There is comparatively low uncertainty on the slope. Our fit passes through all vertical error bars. The green dashed line is a $1\sigma$ confidence interval on the line of best fit. The slope is negative, implying that higher CO$_2$ abundance is correlated with lower active fractions. The black solid line for the active area has an equation of y=(-0.81$\pm$0.24)x+(19.97$\pm$2.89). Our fit encompasses all error bars. We include the water active area in Fig. \ref{fig:kopffplot}b because the uncertainty on the active fraction shown in Fig. \ref{fig:kopffplot}a is predominantly due to the uncertainty on Kopff's size. The uncertainties given in the line of best fit for Fig. \ref{fig:kopffplot}b are more reflective of the relevant data than the fit for Fig. \ref{fig:kopffplot}a, whose uncertainties are affected by the large uncertainty on Kopff's size.

\begin{deluxetable}{lcc}

\tablecaption{Values for Kopff Related to Active Fraction} \label{tab:kopff}

\tablehead{\colhead{} & \colhead{Pre-Perihelion} & \colhead{Post Perihelion}
} 

\startdata
CO$_2$ Abundance ($\%$) & 19.3$\pm$1.9 & 7.4$\pm$0.7 \\
Water Active Area (km$^2$) & 5.1$\pm$0.4 & 14.1$\pm$0.9 \\
Water Active Fraction ($\%$) & 16.1$\pm$9.4 & 44.5$\pm$25.9 \\ 
CO$_2$ Active Area (km$^2$) & 0.286$\pm$0.022 & 0.092$\pm$0.006\\
CO$_2$ Active Fraction ($\%$) & 0.9$\pm$0.5 & 0.3$\pm$0.2
\enddata

\end{deluxetable}

Table \ref{tab:kopff} presents the water and CO$_2$ active areas, and water and CO$_2$ active fractions, as well as the CO$_2$ abundance for 22P/Kopff. We calculated the CO$_2$ active area using a similar method to our earlier calculations of H$_2$O active area, using the \citet{CowanAHearn1979} sublimation model. Looking at the active fractions pre- and post- perihelion, the water active fraction increases and the CO$_2$ active fraction decreases slightly. However, the uncertainties on the active fraction are dominated by the uncertainty on the surface area of Kopff itself. Instead, we examine the active areas, whose associated uncertainties show the changes more clearly. Since we are comparing two time periods for the same comet instead of comparing across comets, the active area is a more effective means to track changes pre- and post-perihelion. The water active area increases, while the CO$_2$ active area decreases. We see increasing water production that is not associated with a corresponding increase in CO$_2$ production. This observation indicates that the new active areas that start to sublimate may not be CO$_2$ rich. The comet outgassed more water and less CO$_2$ post-perihelion compared to pre-perihelion. Due to those trends, the overall CO$_2$ abundance decreases. It is difficult, however, to generalize these trends to all comets since they could be unique to Kopff. 

The negative slopes in Fig. \ref{fig:kopffplot} also match with the overall lack of correlation in Fig. \ref{fig:cometsplot}. Some other factor(s) are causing the differing water active area and CO$_2$ active area rates, aside from the overall CO$_2$ abundance. Whether these factors could be related to thermal evolution due to repeated perihelion passages is an intriguing question, but investigating it further is beyond the scope of this work.

\section{Conclusions} \label{sec:conclu}
In this work, we presented a CO$_2$ abundance measurement for comet HMP during its 2017 apparition using the forbidden oxygen ratio as a ground-based proxy for space-based observations. We found an [\ion{O}{1}] line ratio of 0.077$\pm$0.004. Using release rates B, we report a preferred value for the CO$_2$ to H$_2$O ratio of $22.9\pm1.4\%$. Release rates A yield a ratio of $15.2\pm0.9\%$. 

Our water production rate of $(2.14\pm0.16)\times10^{27}$ mol s$^{-1}$ is similar to those found by \citet{combi2020} and~\cite{DelloRusso2020}, but discrepancies between our results and those of~\cite{combi2020} may be due to an icy grain source in the coma. When we compare HMP to Wirtanen and Hartley 2, we see similarly high CO$_2$, which, in isolation, could support the icy grain hypothesis.

However, when we take the broader context of eight other comets' water active fraction compared to their CO$_2$ abundances, the data do not support the hypothesis that high CO$_2$ abundances are required to drive out the icy grains. We found no correlation between active fraction and CO$_2$ abundance. If the icy grain hypothesis were accurate, we would expect a statistically significant positive correlation. We do not see that correlation. Thus, there are other factors that could lead to a high active fraction, not just CO$_2$ abundance. 

When we perform a case study of 22P/Kopff, we see an overall negative correlation between active fraction and CO$_2$ abundance. Comparing post-perihelion and pre-perihelion observations, we see an increase in the water active area and a slight decrease in the CO$_2$ active area. These trends lead to an overall higher water active fraction after Kopff's perihelion passage. Thus, there may be some complicating factors that cause more water to sublimate without a corresponding increase in CO$_2$ active fraction.

The source of hyperactivity in comets remains an open question, as our results suggest that it is not always the case that high rates of CO$_2$ sublimation will result in the lifting of icy grains off the surface that causes higher water active fraction. High CO$_2$ abundance may be necessary but not sufficient for hyperactivity. There are complicating factors to be studied. Is there another method that leads to hyperactivity aside from CO$_2$ sublimation? What conditions are necessary for hyperactivity? Is there a heliocentric distance or thermal evolution factor that contributes to these trends? We suggest that there are additional, yet-unknown factors that affect hyperactivity, outside of high CO$_2$ abundance. 

\begin{acknowledgments}
We would like to thank the two anonymous reviewers for their helpful comments which improved the quality of this manuscript. We extend our thanks to the NASA Office of STEM Engagement Internship Program, which provided funding for the work presented. We would also like to acknowledge the JPL Horizons System, which was used to generate ephemerides for comet calculations. 
\end{acknowledgments}

\vspace{5mm}
\facilities{McDonald Observatory}

\software{pyspeckit (https://pyspeckit.readthedocs.io), \newline QPINT1D (https://www.nv5geospatialsoftware.com/docs/qpint1d.html)
          }

\bibliography{references}{}
\bibliographystyle{aasjournal}

\end{document}